\documentclass[11pt,journal,onecolumn]{IEEEtran}

\usepackage{amsmath,url}
\usepackage{amssymb,amsbsy,verbatim,array}
\usepackage{epsfig,pstricks,psfrag,theorem,cite,graphicx,xcomment,subfigure}

\let\intern=\iffalse

\def\figref#1{Fig.\,\ref{#1}}%
\def\E{\mathbb{E}}

\def\R{\mathbb{R}}

\def\ie{{\em i.e.}}

\def\dd{\mathrm{d}}

\def\one{\mathbf{1}}

\def\nucl{\nu_{\rm cl}}
\def\nup{\nu_{\rm p}}

\newtheorem{fact}{Fact}



\let\figs=\iftrue

\begin{document}
\title{A Versatile Dependent Model for\\Heterogeneous Cellular Networks}
\author{Martin Haenggi\\
University of Notre Dame\\
\today}

\maketitle
\begin{abstract}
We propose a new model for heterogeneous cellular networks that incorporates dependencies between
the layers. In particular, it places lower-tier base stations at locations that are poorly covered by
the macrocells, and it includes a small-cell model for the case where the goal is to enhance network capacity.
\end{abstract}
\section{Motivation}
Due to the increasing spatial irregularity of cellular systems, point process models are ideally suited for analysis and simulation.
When modeling multiple tiers, including macro-, micro-, and femto-tiers, a natural choice is the superposition of independent Poisson point processes (PPPs), each one modeling the locations of the base stations of a particular tier. While this model has obvious analytical advantages, it does not accurately capture the deployment of small cells that are deployed with the
objective of either enhancing the {\em coverage} or the {\em capacity} of the network.

As a consequence, it appears more realistic to use a model that incorporates such dependencies between the tiers
and within the tiers. At the same time, the analytical tractability of the independent multi-Poisson model should not be completely lost. Here we propose a new model that provides a trade-off between accuracy and tractability.

\section{The Proposed Model}
\begin{figure}
\subfigure[BS locations of the four tiers and Voronoi edges of tier 1. The small cells of tier 2 are restricted to these edges, whereas
those of tier 3 are located at the Voronoi vertices.]{\epsfig{file=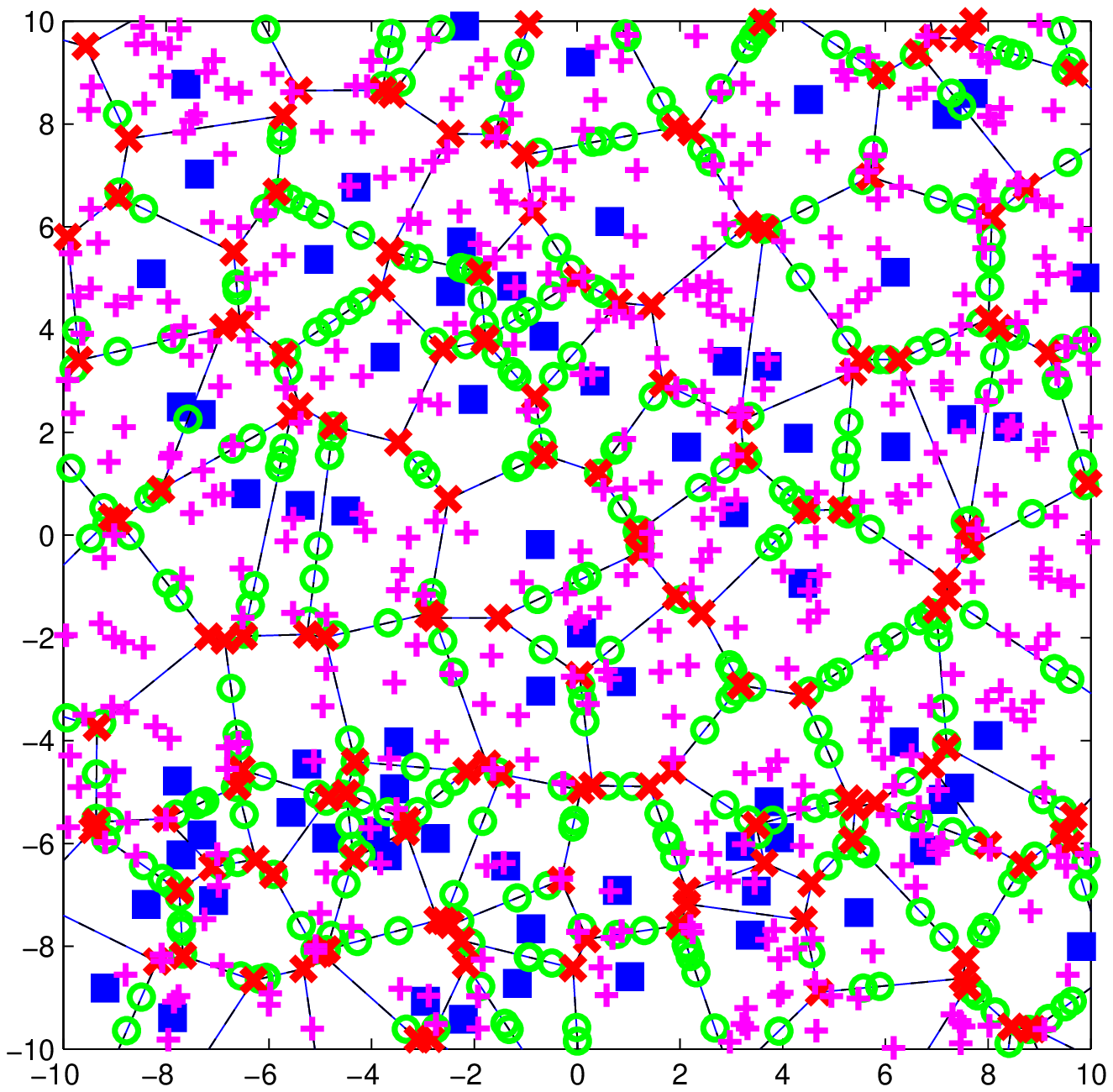,width=.47\columnwidth}}\hfill
\subfigure[BS locations and associated cells.]{\epsfig{file=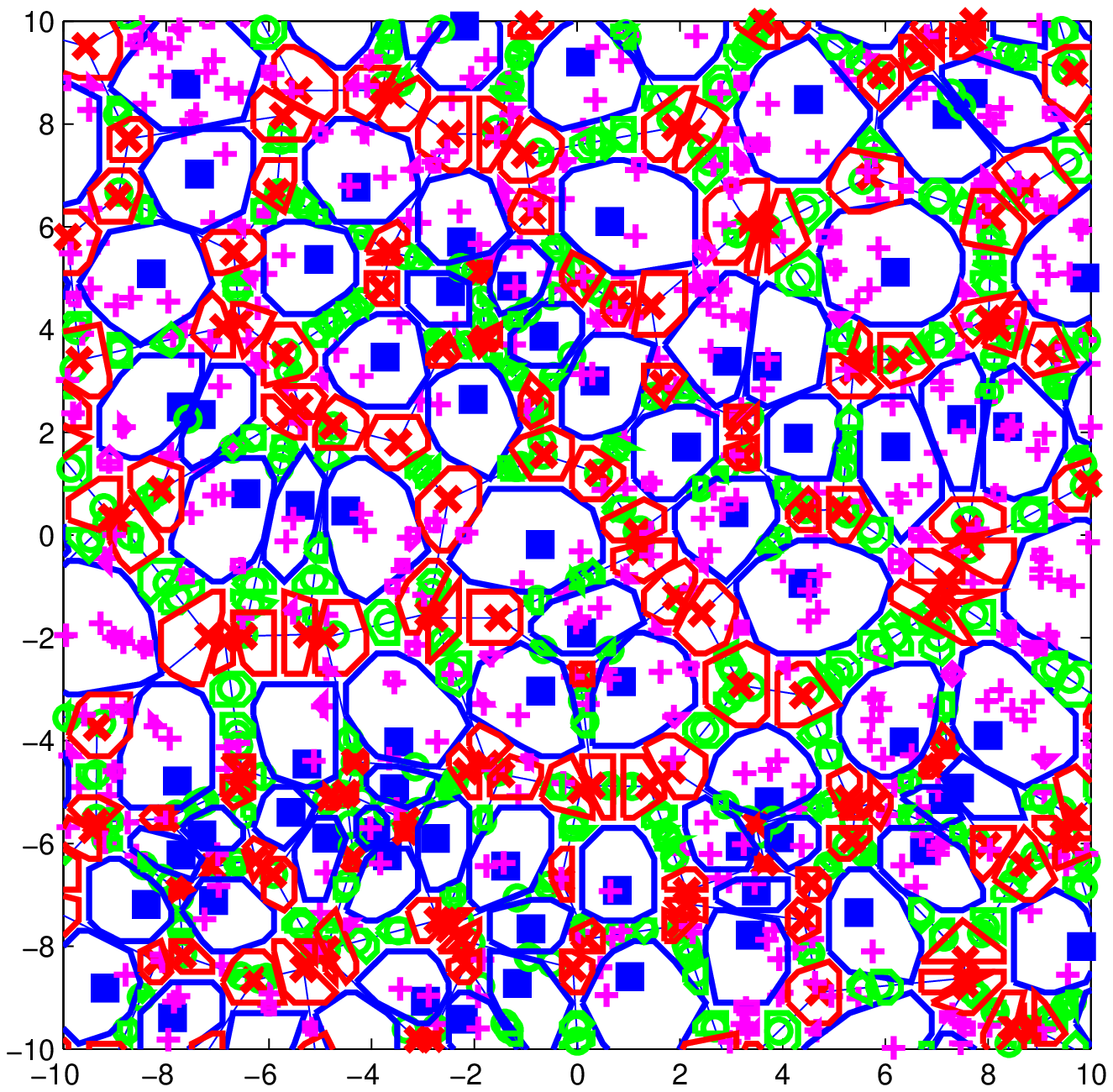,width=.47\columnwidth}}\\
\subfigure[RSS contour plot (dB).]{\epsfig{file=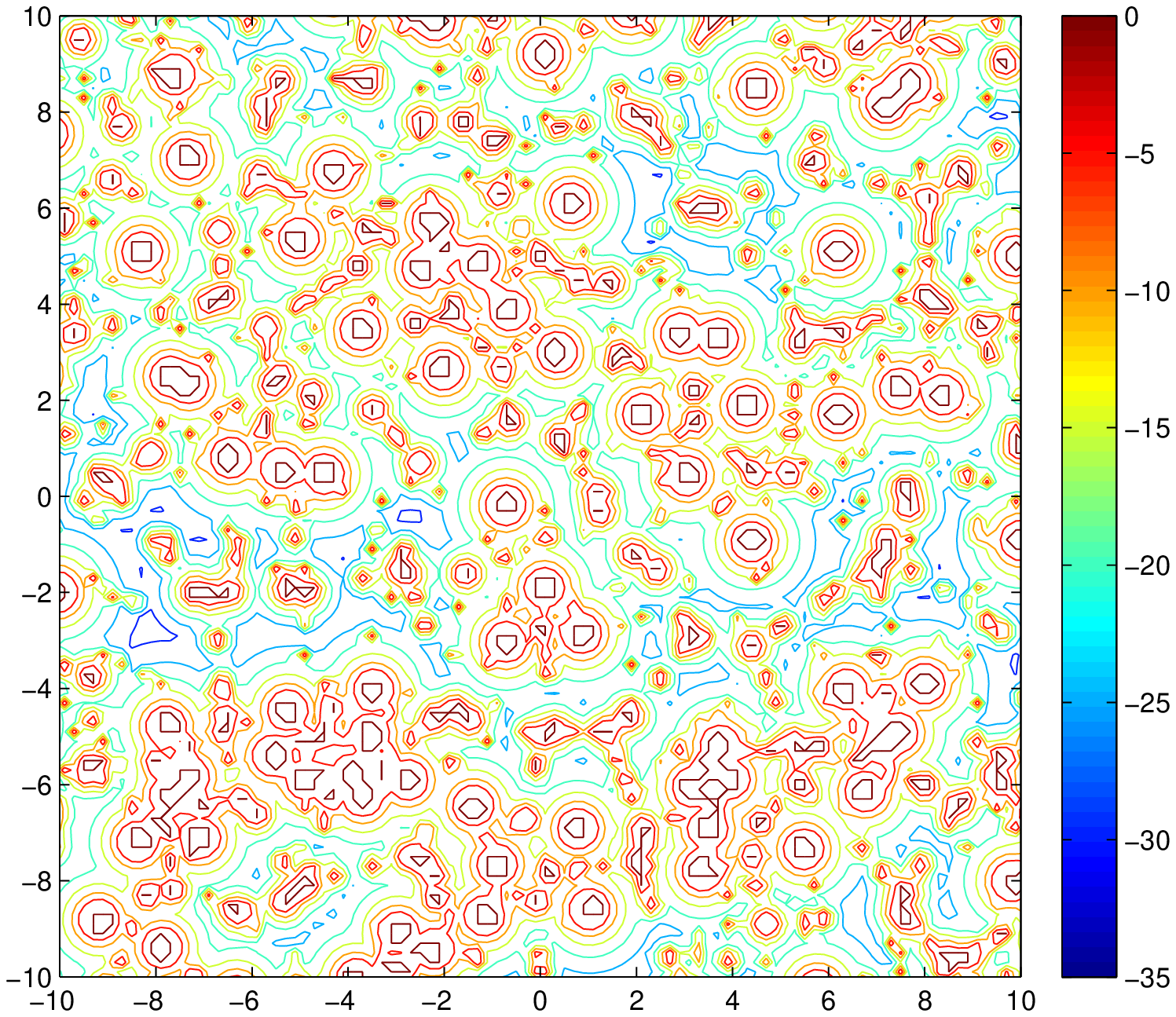,width=.52\columnwidth,height=.46\columnwidth}}\hfill
\subfigure[Coverage by tier at -30 dB. The white region (1.5\% of the area) is not covered at this signal strength.]{\epsfig{file=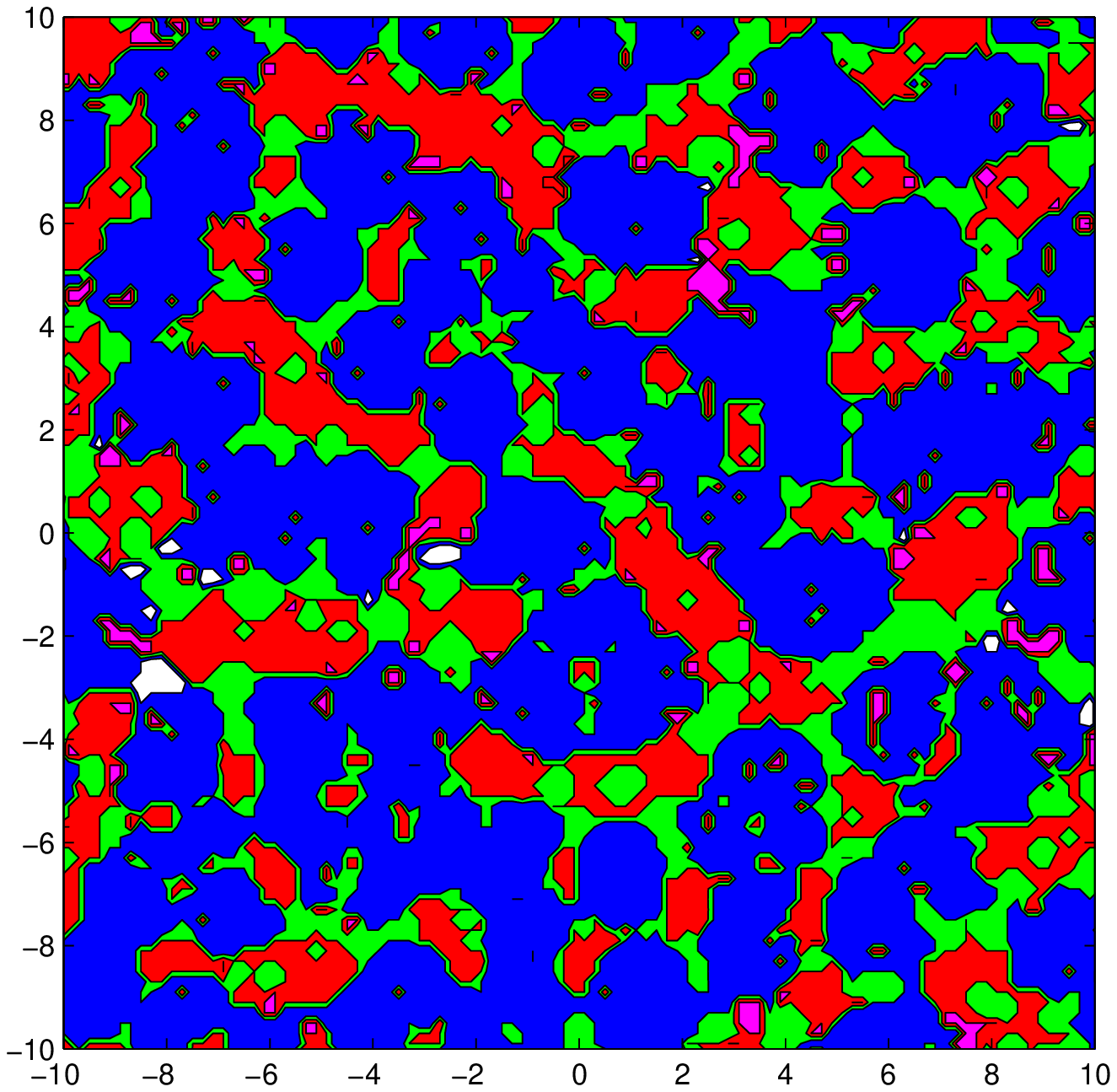,width=.47\columnwidth}}\\
\caption{A complete HetNet with $\lambda=2/10$, $\mu=1$, $p=1$, $\nu=1$ and power levels of 50, 30, 40, and 20 dBm. Tiers 1 through 4 are formed by the blue squares, green circles, red 'x's, and magenta '+'s,
respectively. The total (theoretical) density is 2.49, and the total power is 9 kW. The uncovered area fraction is 0.9\%.}
\label{fig:hetnet1}
\end{figure}

The proposed model consists of four tiers. In its basic form, the tiers are defined as follows.
\begin{enumerate}
\item Tier 1 consists of a homogeneous PPP of intensity $\lambda$ on the plane.
\item Tier 2 consists of a non-homogeneous PPP that is restricted to the edges of the Voronoi cells
of tier 1. On each Voronoi edge, a PPP of intensity $\mu$ (points per unit length) is placed.
\item Tier 3 consists of an independent thinning of the Voronoi vertices\footnote{The locations where three Voronoi edges meet.} of tier 1 with retaining probability $p$. If $p=1$, all the Voronoi vertices are retained.
\item Tier 4 is again a homogeneous PPP of intensity $\nu$ on the plane.
\end{enumerate}
The reason for modeling tiers 2 and 3 in this manner is that Voronoi edges and vertices comprise the locations with poorest coverage by the macro-tier 1. Points on the Voronoi edges have equal distance to two macro-BSs, while the Voronoi vertices have equal distance to three macro-BSs. These locations are thus
natural choices for the placement of additional BSs to improve coverage.
Tier 4 can be used to model femtocell BSs. An example of a realization of such a HetNet model is shown in
\figref{fig:hetnet1}(a). To illustrate the coverage properties, we assign a transmit power level to the BSs in each
tier and use a simple path loss model of the form $(d/d_0)^{-\alpha}$ with $d_0=0.01$.
\figref{fig:hetnet1}(b) visualizes the cells associated to each transmission point
 (which include the locations of the plane that receive the strongest signal from each transmission point),
(c) shows a contour plot of the received signal strength (RSS), and (d) shows a coverage map, where each
location that is covered at at least -30 dB is colored in the color of the respective tier, while uncovered locations
are left white.

\begin{figure}
\subfigure[BS locations (tiers 1 and 4) and Voronoi edges of tier 1.]{\epsfig{file=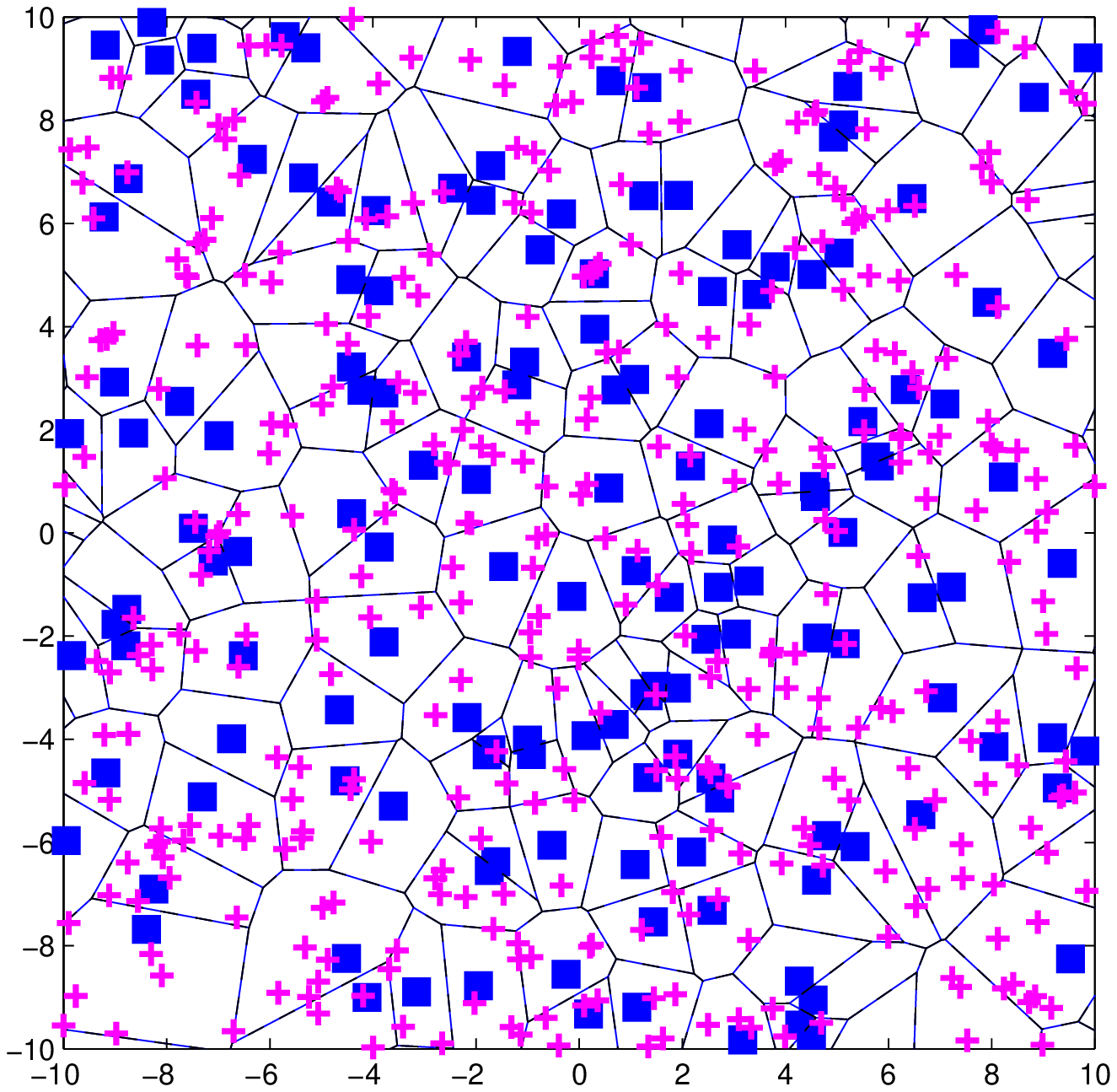,width=.47\columnwidth}}\hfill
\subfigure[BS locations and associated cells.]{\epsfig{file=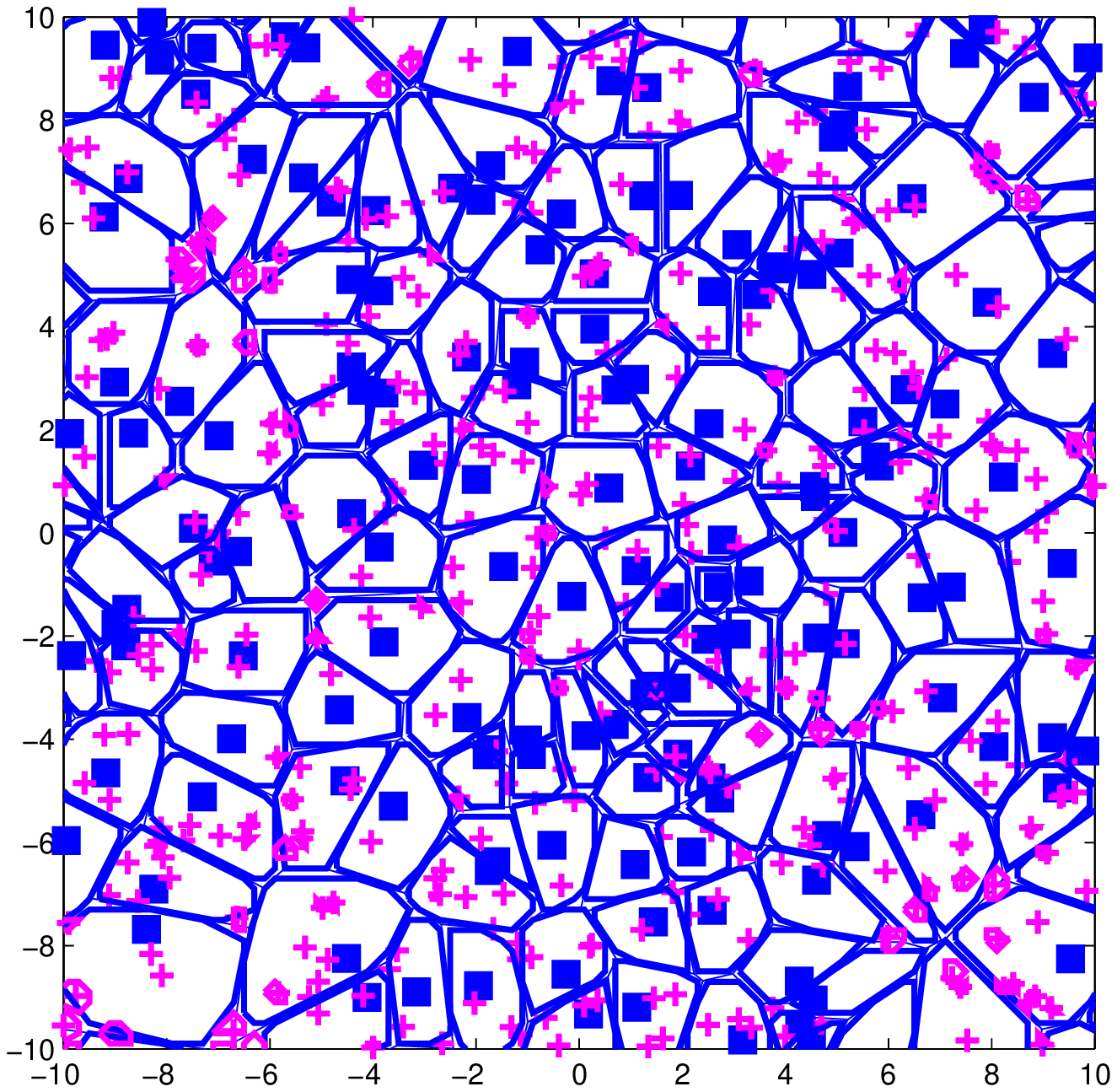,width=.47\columnwidth}}\\
\subfigure[RSS contour plot (dB).]{\epsfig{file=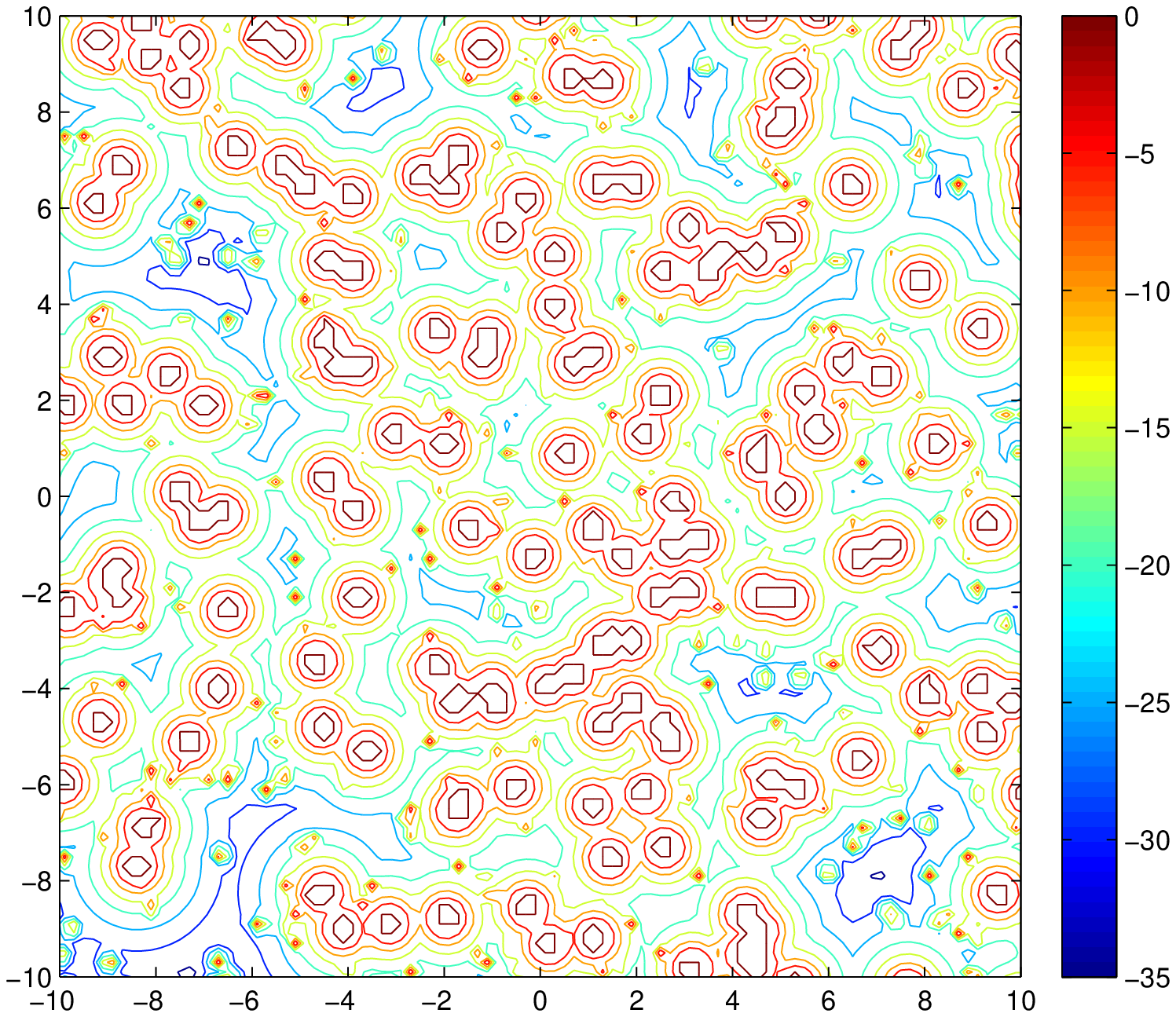,width=.52\columnwidth,height=.46\columnwidth}}\hfill
\subfigure[Coverage by tier at -30 dB (blue for tier 1 and magenta for tier 4). The white region (2.6\% of the area) is not covered at this signal strength.]{\epsfig{file=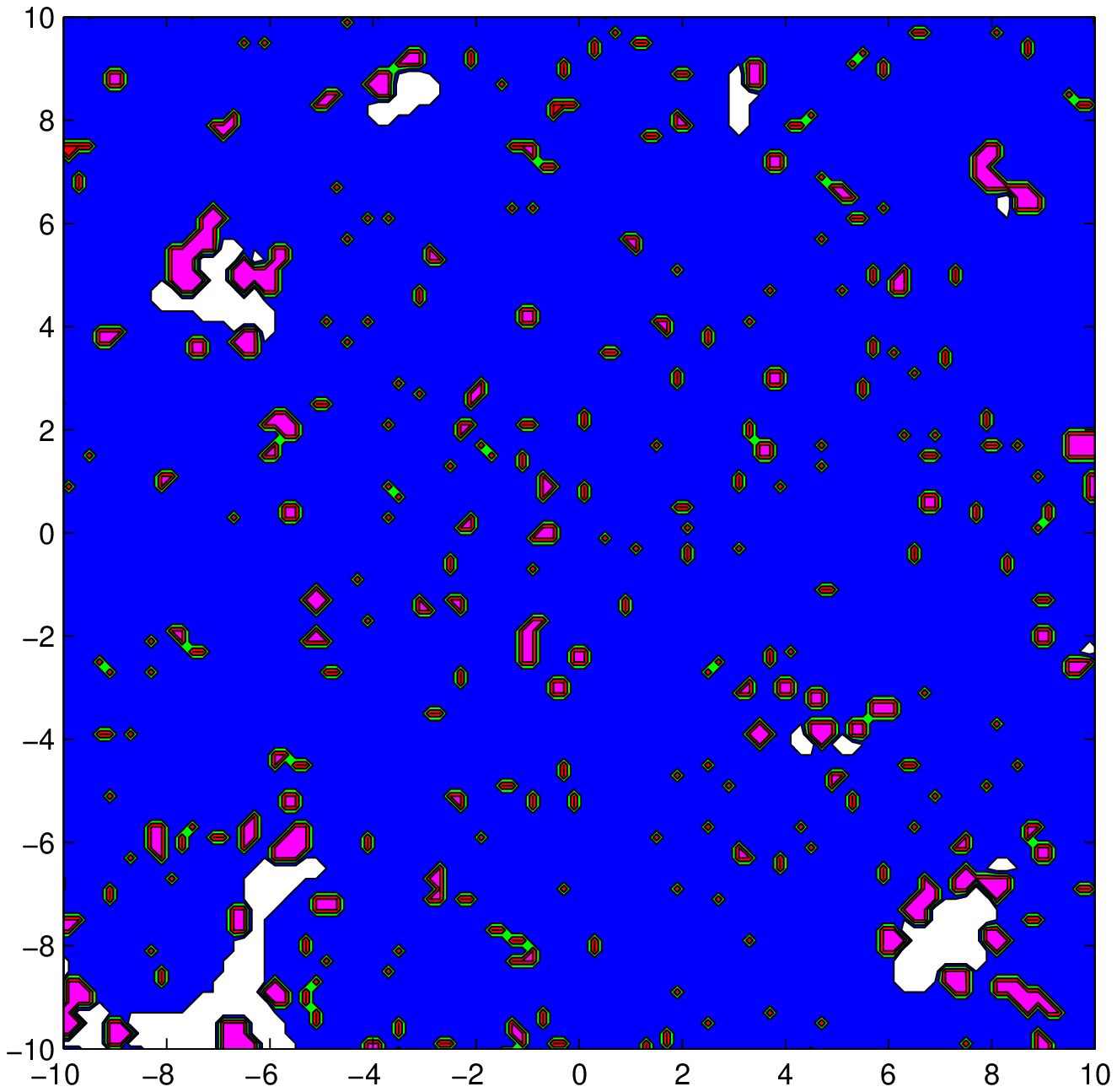,width=.47\columnwidth}}\\
\caption{A two-tier HetNet with $\lambda=4/10$, $\mu=0$, $p=0$, $\nu=1$ and power levels of 50 and 20 dBm. Tiers 1 and 4 are formed by the blue squares and and magenta '+'s,
respectively. The total (theoretical) density is 1.4, and the total power is 15 kW. The uncovered area fraction is 2.6 \%.}
\label{fig:hetnet1b}
\end{figure}

In contrast, \figref{fig:hetnet1b} shows a two-tier HetNet that is comprised of only independent elements.
The uncovered fraction is almost tripled, although the network consumes 66\% more power (15 kW vs.~9 kW).

\figs
\fi

\figs
\fi


\begin{fact}
Each tier is a stationary point process, and the intensities are $\lambda$, $2\mu\sqrt\lambda$, $2p\lambda$, and $\nu$, respectively.
\end{fact}
The only non-obvious intensity is perhaps the one for tier 2. It follows from the
expected length of the Voronoi cell perimeter, which is $4/\sqrt\lambda$.
Fact 1 also holds if tier 1 is a general stationary point processes---as long as the points are in {\em general quadratic position}, which means that no three points lie on a line and no four points lie on a circle.

\section{Special Cases}
\begin{figure}
\subfigure[Base station locations of tiers 1 and 3 and associated cells.]{\epsfig{file=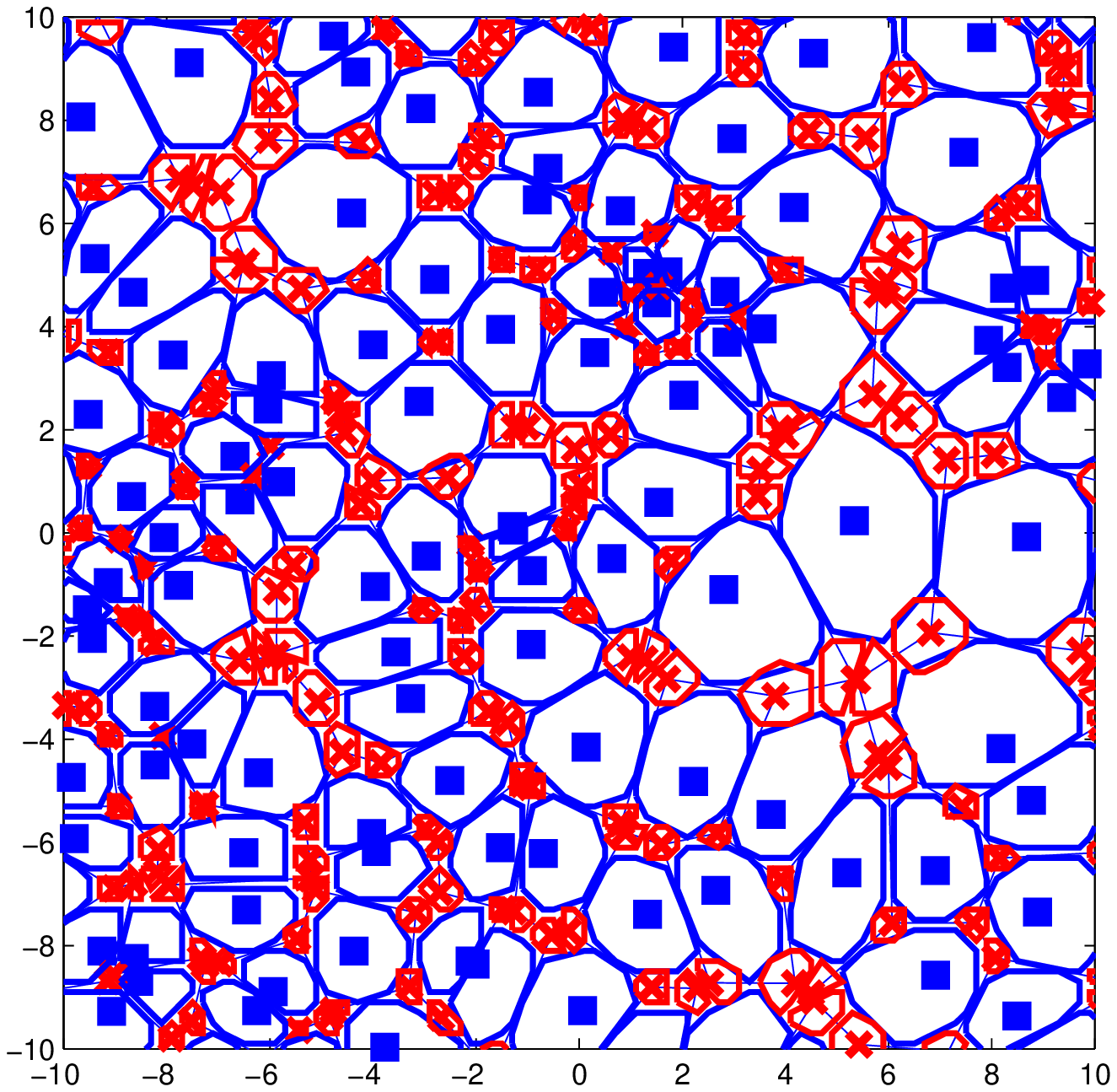,width=.48\columnwidth}}\hfill
\subfigure[Coverage map at -30 dB.  1.7\% of the square region is not covered.]{\epsfig{file=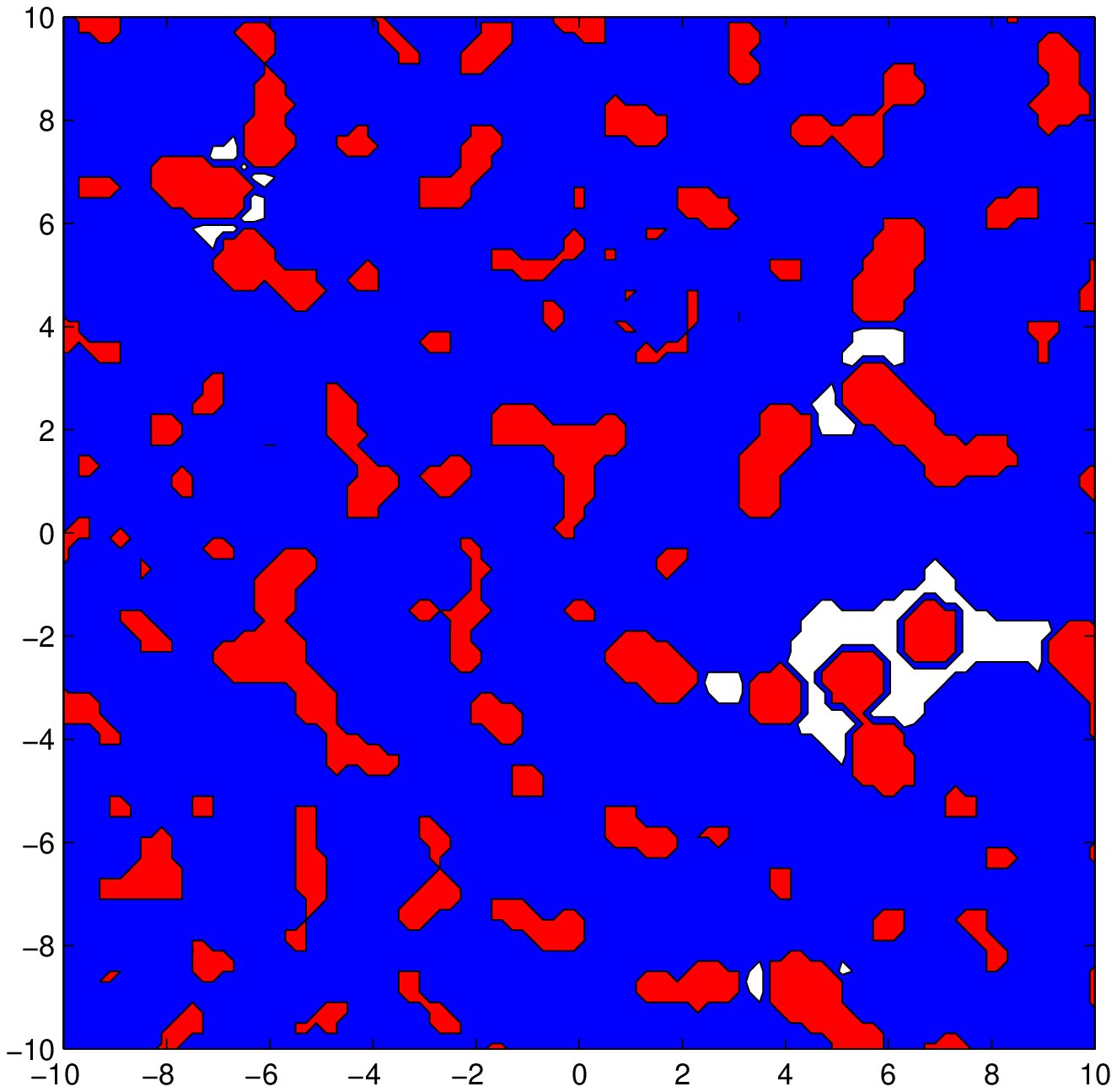,width=.48\columnwidth}}\\
\caption{A HetNet consisting of Tiers 1 and 3 with $\lambda=1/4$, $\mu=0$, $p=1$, $\nu=0$ and power levels of 50 and 33 dBm. Tiers 1 and 3 are formed by the blue squares and red 'x's,
respectively. The thin blue lines represent the Voronoi edges. Tier 3 has twice the intensity and uses $1/50$ of the power at each transmission point. It provides a higher signal strength than tier 1 in 22\% of the area. The total power is 10 kW.}
\label{fig:hetnet2}
\end{figure}

The model comprises several special cases of interest, for example:
\begin{enumerate}
\item If $\mu=0$ and $p=0$, the model reduces to the independent two-tier Poisson model, which
has full analytical tractability.
\item If $\mu=0$, $p>0$, and $\nu=0$, it represents the Poisson model with additional BSs at the
locations of the weakest coverage.
\end{enumerate}
Case 2 is of interest to address the question of the coverage gain achievable with small cells placed
at strategic positions. In a homogeneous Poisson cellular network with density $1/4$ with 50 dBm transmit power,
the covered area
fraction is only about 90\% (for the path loss model given above and a threshold of -30 dB).
If small cells with 33 dBm transmit power are placed at all the Voronoi vertices, the total power consumption
changes only slightly by 4\%, but the covered area fraction increases to about 98\%, as shown in \figref{fig:hetnet2}.



\figs
\fi

%


\section{Refinements}
Dependencies between base station locations may also be introduced within each tier. 
Two important cases are discussed here: (1) Imposing a minimum spacing between base stations for more
realistic modeling (Subs. A); (2) Using a cluster model at tier 4 to model the situation where
small cells are placed to enhance the system capacity in regions of high population density (Subs. B).


\subsection{Hard- or soft-core models}
\begin{figure}
\subfigure[BS locations (tiers 1 and 2) and Voronoi edges of tier 1. The small cells of tier 2 are restricted to these edges.]{\epsfig{file=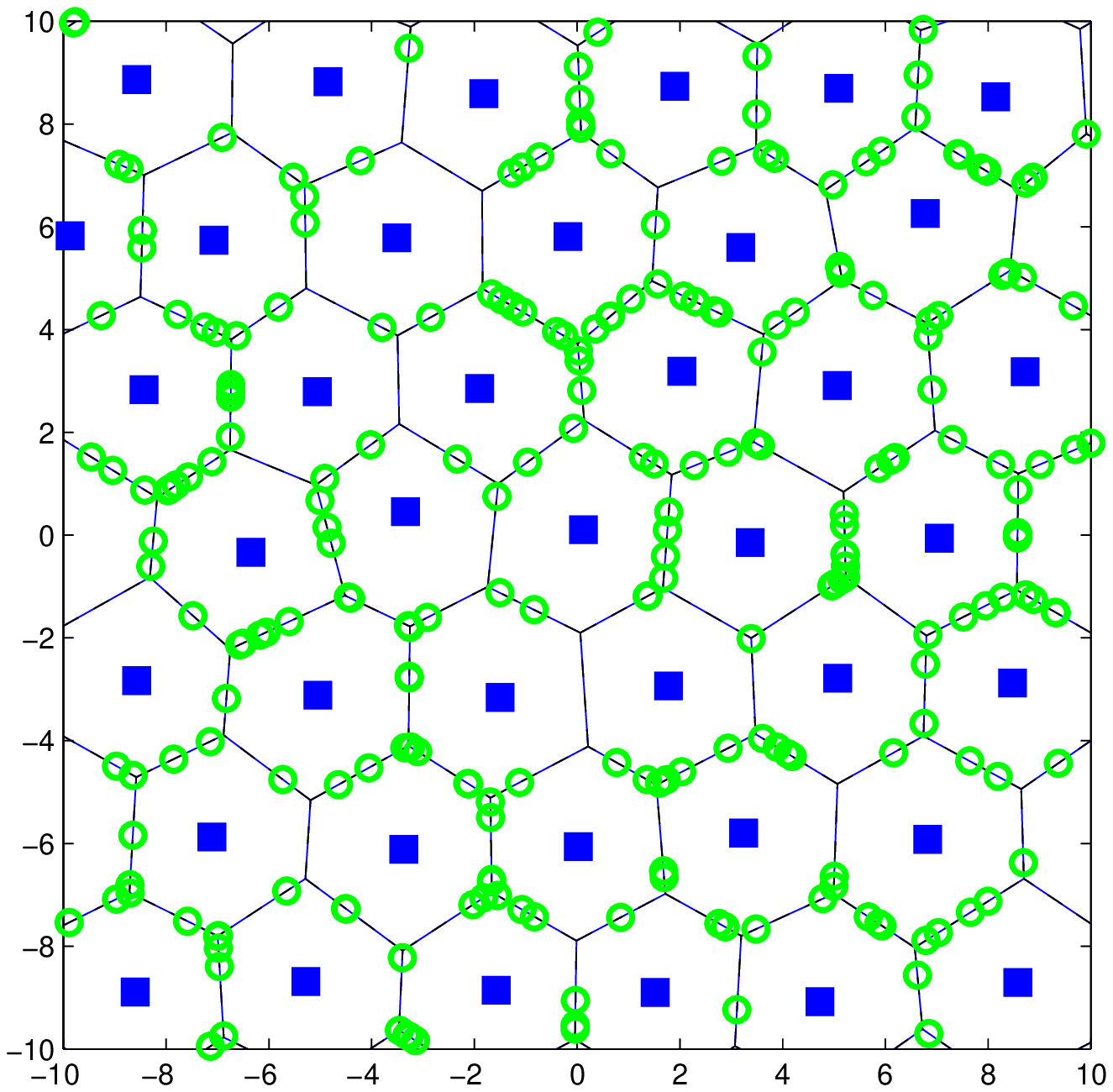,width=.47\columnwidth}}\hfill
\subfigure[BS locations and associated cells.]{\epsfig{file=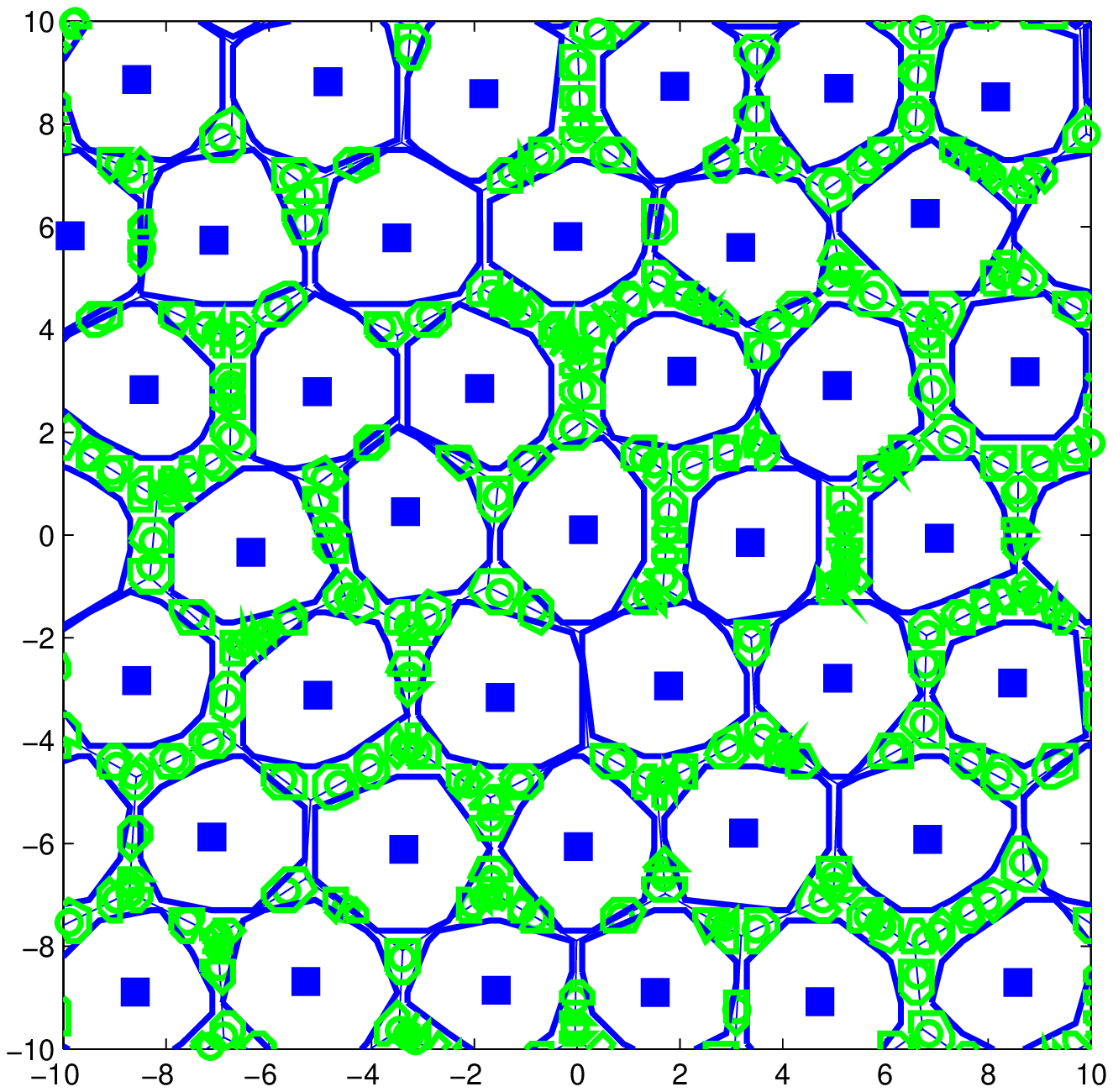,width=.47\columnwidth}}\\
\subfigure[RSS contour plot (dB).]{\epsfig{file=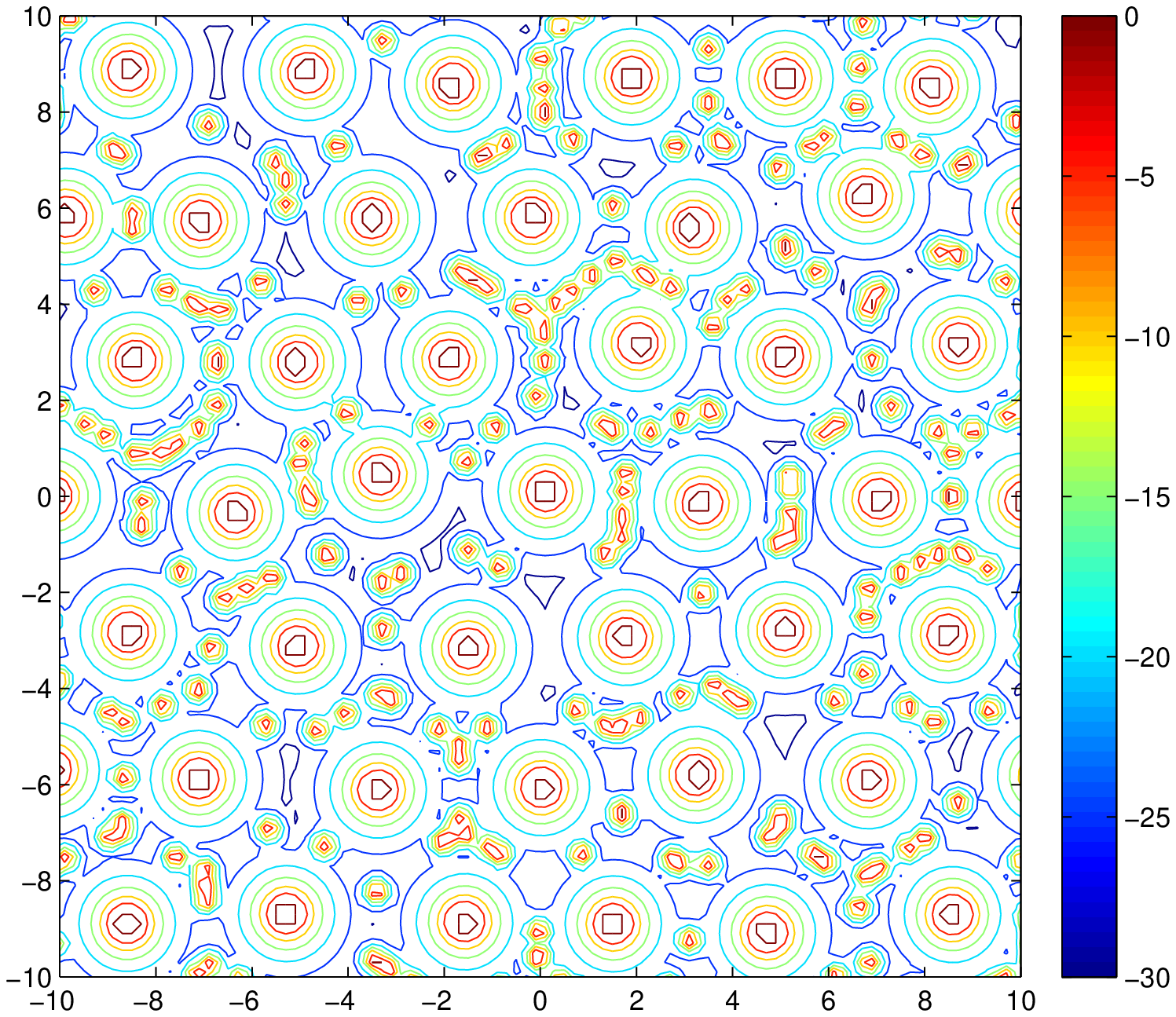,width=.52\columnwidth,height=.46\columnwidth}}\hfill
\subfigure[Coverage by tier at -30 dB (blue for tier 1, green for tier 2). The white region (1.2\% of the area) is not covered at this signal strength.]{\epsfig{file=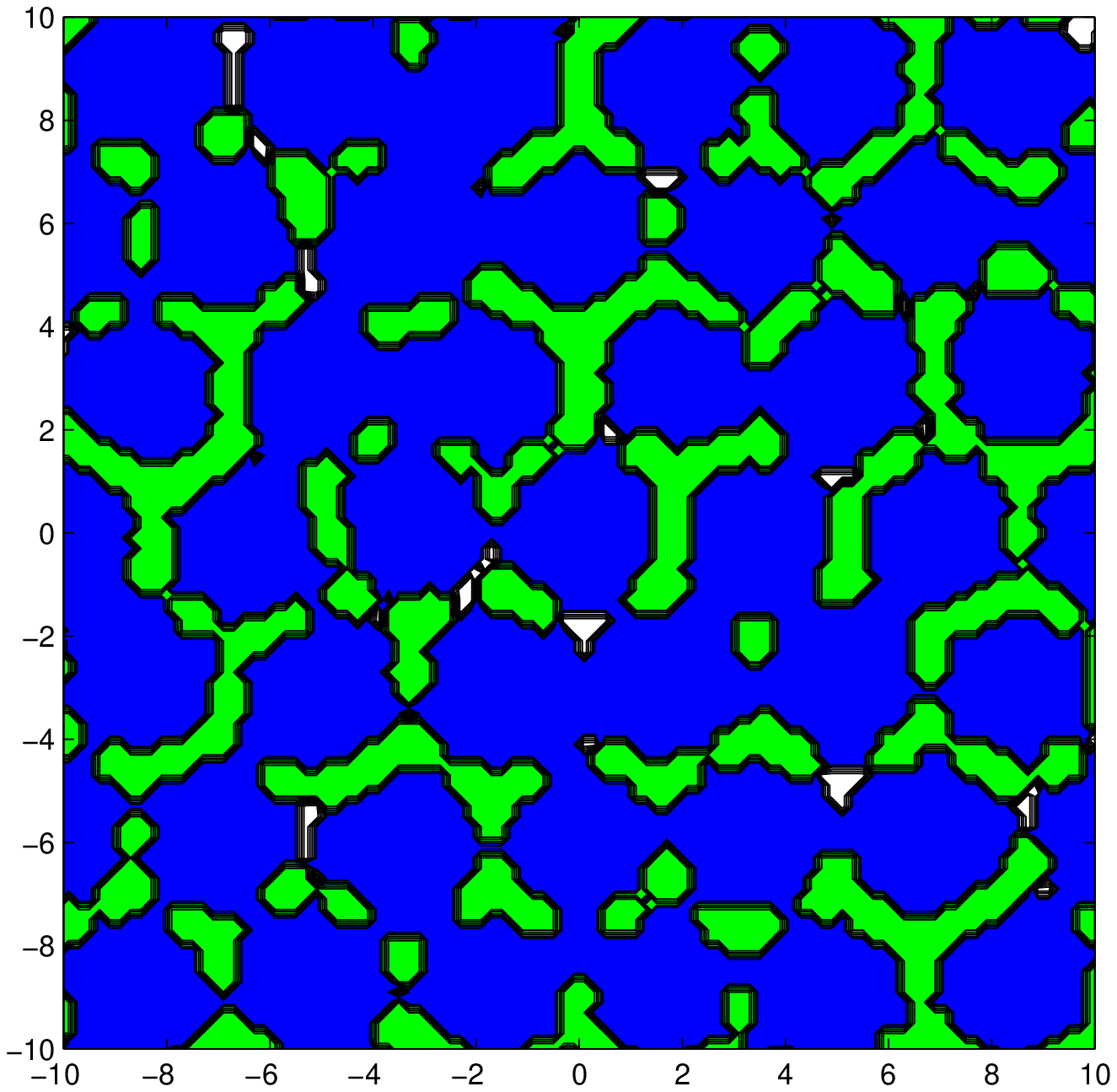,width=.47\columnwidth}}\\
\caption{A two-tier HetNet with $\lambda=1/10$, $\mu=1$, $p=0$, $\nu=0$ and power levels of 50 and 30 dBm. Tiers 1 and 2 are formed by the blue squares and and green circles,
respectively. Tier 1 forms a perturbed triangular lattice (variance 0.04). The total (theoretical) density is 0.73, and the total power is 4.2 kW. The uncovered area fraction is 1.2\%.}
\label{fig:hetnet3}
\end{figure}

The PPPs in tiers 1, 2, and 4 can be replaced by hard- or soft-core models. The most important
scenario is probably the one where tier 1 forms a more regular point process. As an example,
\figref{fig:hetnet3} shows a two-tier HetNet with a perturbed triangular lattice (Gaussian perturbation with variance 0.04) at tier 1.
Tier 2 is Poisson (on the Voronoi edges) of intensity $\mu=1$.

\subsection{Clustered models for tier 4}
Clustering in the lower tier that is independent of tier 1 can be employed to model deployment of picocells in
regions of high population density and/or traffic demand, \ie, when the goal is to achieve higher network capacity via offloading to small cells. Generally, homogeneous independent cluster processes are suitable
to model this tier. Among these, for better tractability, either Poisson cluster processes
or, more generally, stationary Cox processes are preferable.
Since the Cox framework is more general, we present here a Cox process that
is equivalent to a particular Poisson cluster process, the so-called
Mat\'ern cluster process.

Let $\Psi$ denote a stationary point process of population centers of intensity $\nup$
and let $\nucl\colon\R^2\to\R^+$ be the intensity function
of the picocell deployment for a population center at the origin $o$. Given $\Psi$, tier 4 forms a non-stationary PPP of intensity
\[ \lambda_{\Psi}(x)=\sum_{y\in\Psi} \nucl(x-y) \]
The (unconditioned) intensity of tier 4 does not depend on $x$ and follows as
\[ \nu=\E\lambda_{\Psi}(x)=\E \sum_{y\in\Psi} \nucl(x-y)=\nup\int_{\R^2} \nucl(x-y)\dd y. \]
In the Mat\'ern cluster process, $\Psi$ is a homogeneous PPP and 
\[ \nucl(x)=M\frac{\one(\|x\|<r)}{\pi r^2} .\]
The parameter $M$ denotes the mean number of picocells per population center.
The overall density is given by $\nu=M\nup$.

\begin{figure}
\subfigure[BS locations and Voronoi edges of tier 1. The small cells in tier 4 form a
Mat\'urn cluster process. The parent points are marked by cyan stars.]{\epsfig{file=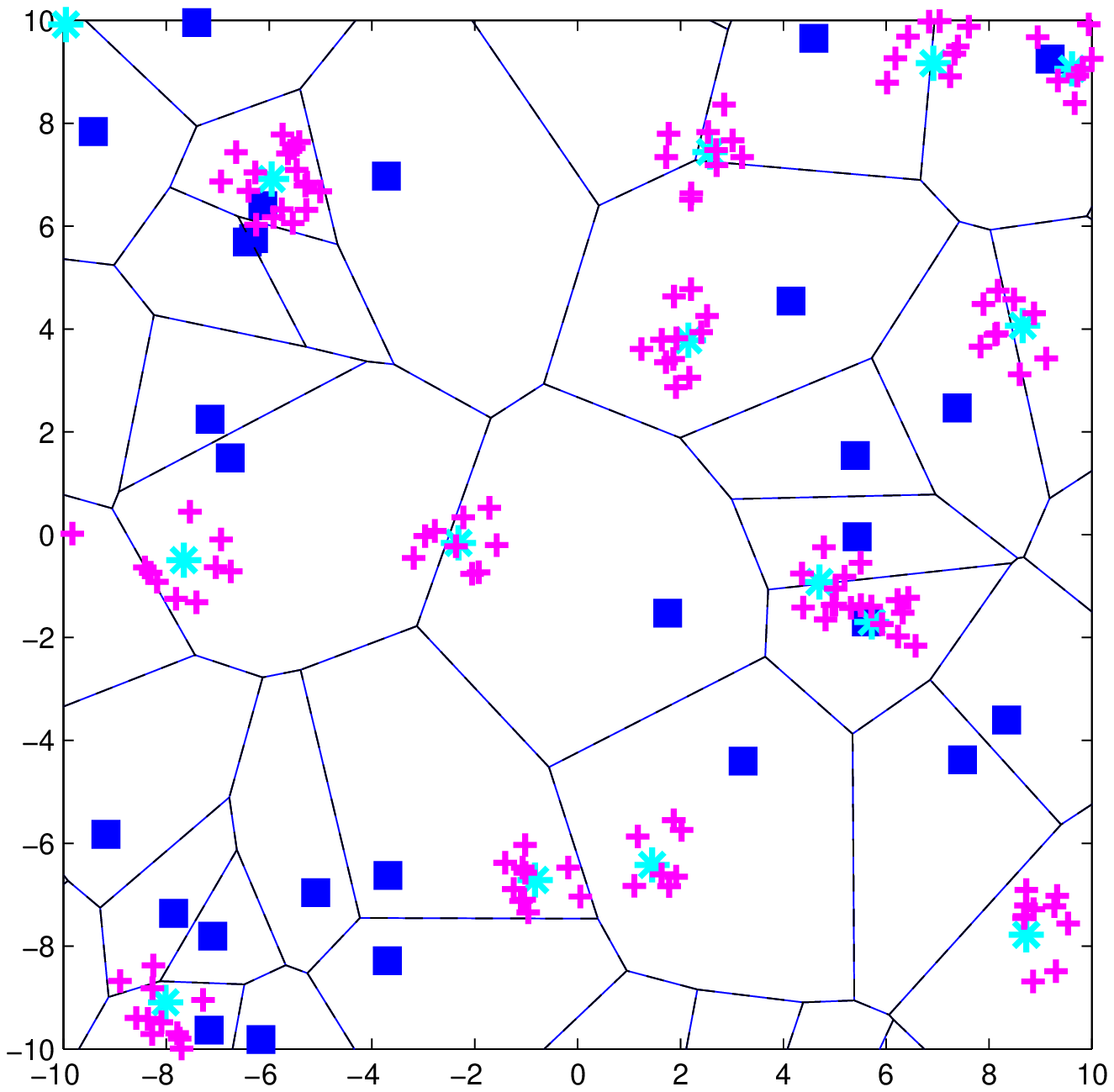,width=.47\columnwidth}}\hfill
\subfigure[BS locations and associated cells.]{\epsfig{file=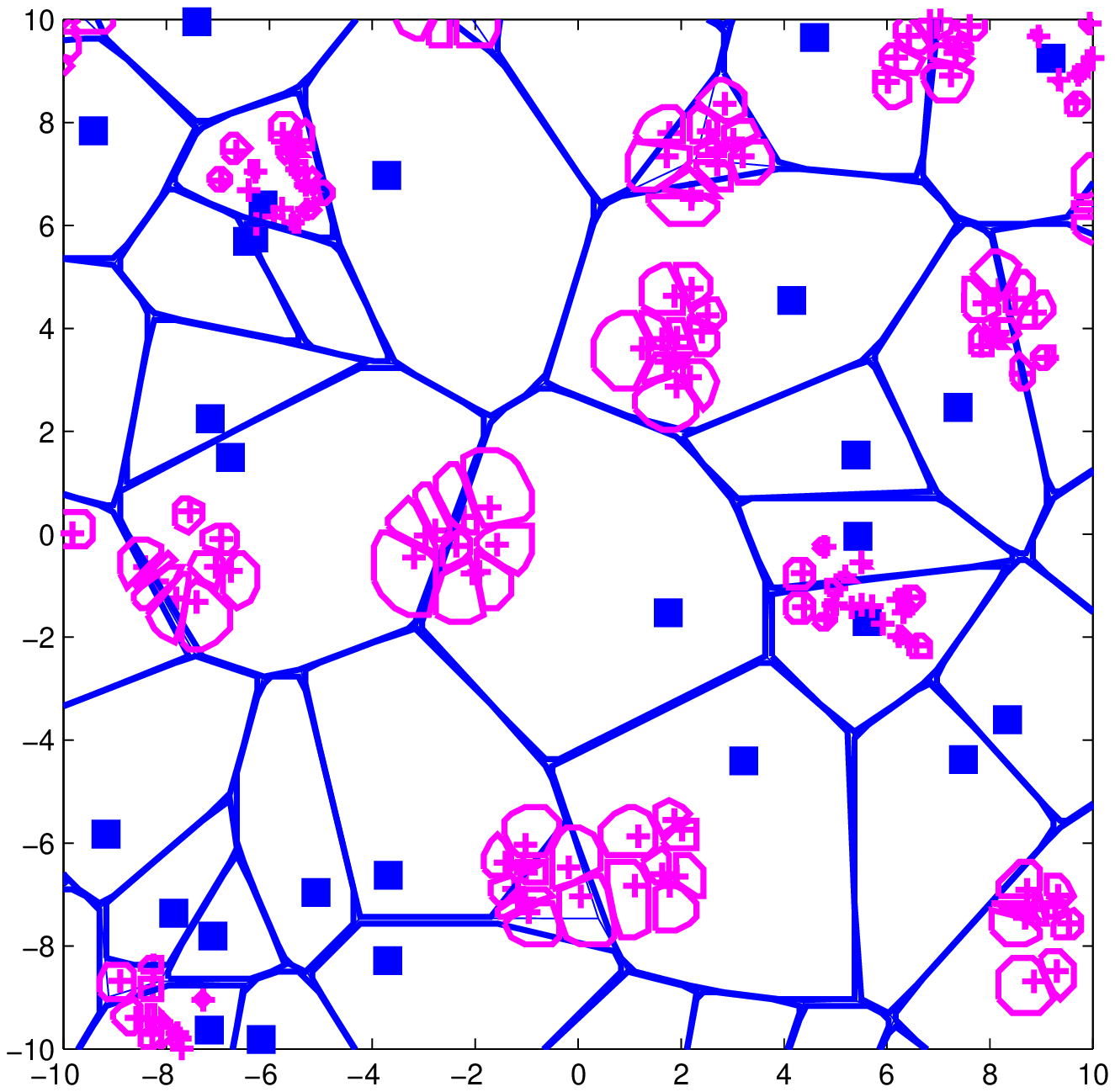,width=.47\columnwidth}}\\
\subfigure[RSS contour plot (dB).]{\epsfig{file=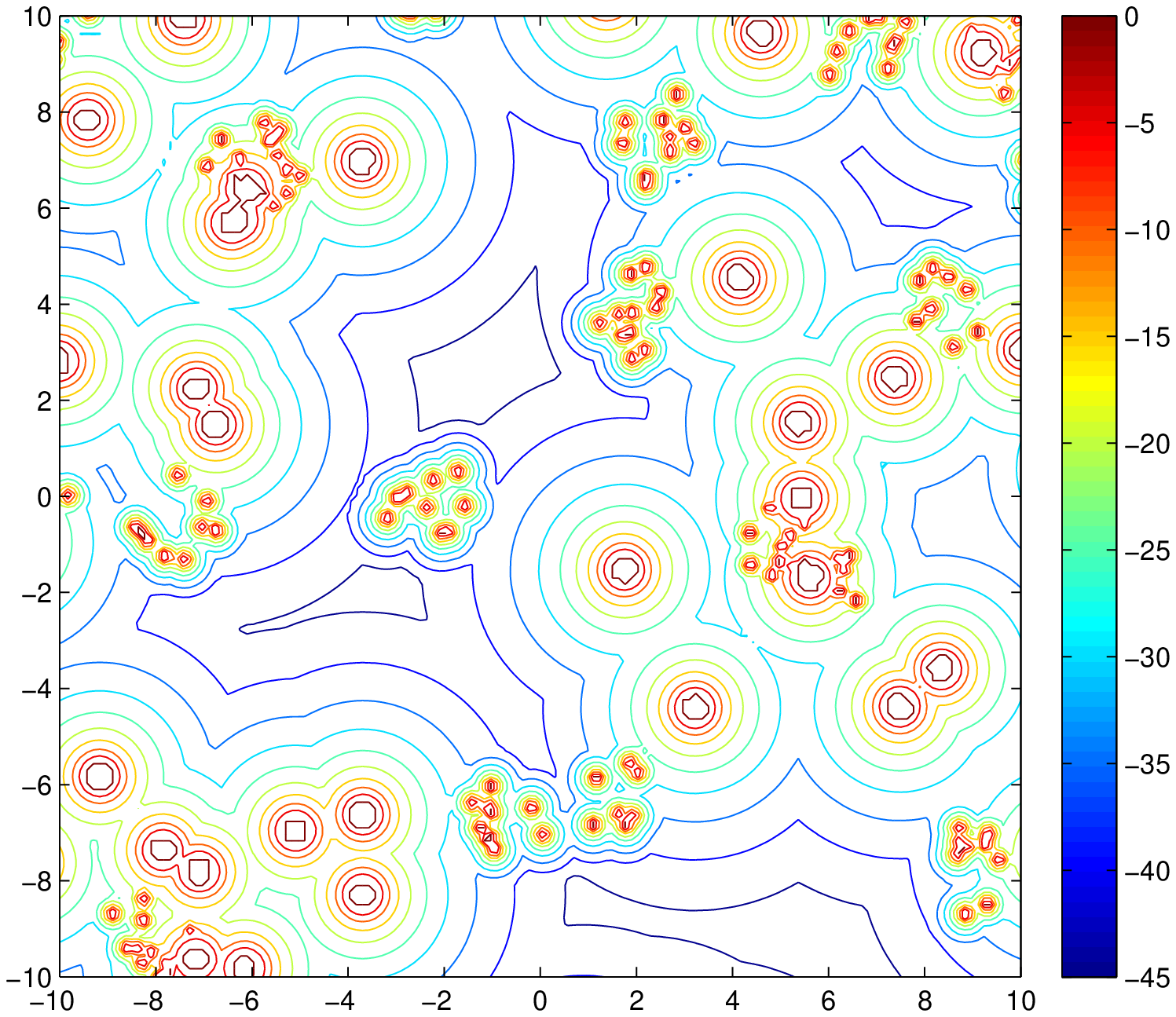,width=.52\columnwidth,height=.46\columnwidth}}\hfill
\subfigure[Cell association (blue for tier 1, magenta for tier 4).]{\epsfig{file=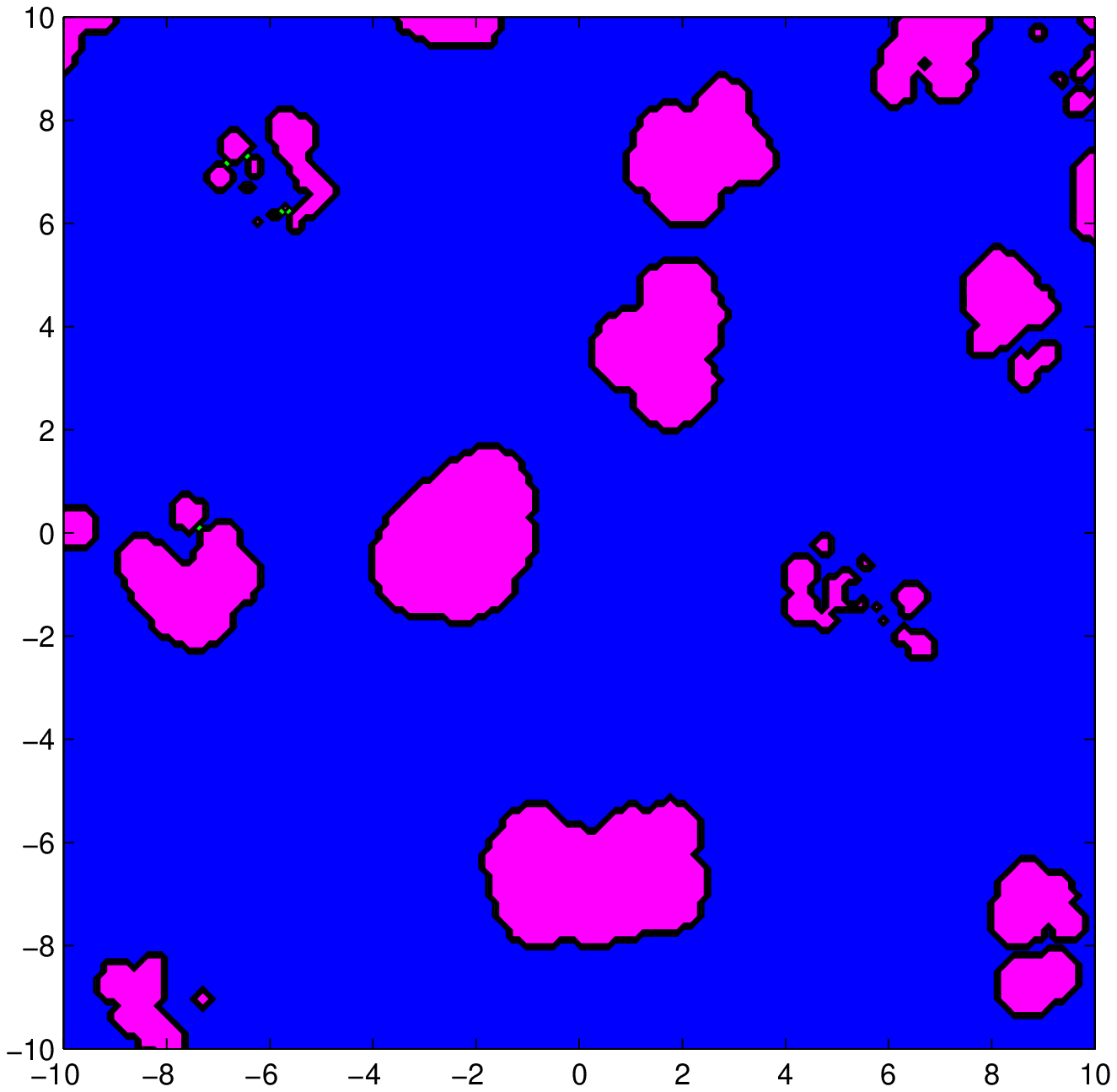,width=.47\columnwidth}}\\
\caption{A two-tier HetNet with $\lambda=1/10$, $\mu=0$, $p=0$, $\nu=1/2$ and power levels of 50 and 26 dBm. Tiers 1 and 4 are formed by the blue squares and and the magenta '+'s,
respectively. Tier 1 forms a PPP, while tier 4 forms a Poisson cluster process with parent density $1/20$, cluster radius $1$, and $M=10$ points per cluster on average. The total (theoretical) density is 0.6. About 20\% of the area receives higher signal strength from tier 4.}
\label{fig:hetnet4}
\end{figure}

\figref{fig:hetnet4} shows a realization of a two-tier HetNet where the lower tier (tier 4 in our model)
is a Mat\'ern cluster process (which is a particular Poisson cluster process and also a Cox process).

\figs
\fi

%
\figs
\fi


\section{Conclusions}
We have introduced a new HetNet model that is versatile enough to model network deployments whose objective is enhanced coverage as well as those whose objective is increased capacity.
The model attempts to strike a balance between analytical tractability and practicality by introducing dependencies across tiers and, in the refined model, within tiers.
It also provides a solid foundation and hopefully common ground for HetNet simulations.

\bibliographystyle{IEEEtr}
\bibliography{net}

\end{document}